# DARE MISSION DESIGN: LOW RFI OBSERVATIONS FROM A LOW-ALTITUDE FROZEN LUNAR ORBIT

Laura Plice[*], Ken Galal[†], and Jack O. Burns[‡]

The Dark Ages Radio Experiment (DARE) seeks to study the cosmic Dark Ages approximately 80 to 420 million years after the Big Bang. Observations require truly quiet radio conditions, shielded from Sun and Earth electromagnetic (EM) emissions, on the far side of the Moon. DARE's science orbit is a frozen orbit with respect to lunar gravitational perturbations. The altitude and orientation of the orbit remain nearly fixed indefinitely, maximizing science time without the need for maintenance. DARE's observation targets avoid the galactic center and enable investigation of the universe's first stars and galaxies.

## INTRODUCTION

The Dark Ages Radio Experiment (DARE), currently proposed for NASA's 2016 Astrophysics Medium Explorer (MIDEX) announcement of opportunity with a launch date in August 2023, seeks to study the universe in the cosmic Dark Ages approximately 80 to 420 million years after the Big Bang.[1,2] DARE's observations require conditions truly free of radio frequency interference (RFI), shielded from Sun and Earth EM emissions, a zone which is only available on the far side of the Moon. To maximize time in the best science conditions, DARE requires a low, equatorial lunar orbit.

For DARE, diffraction effects define quiet cones in the anti-Earth and anti-Sun sides of the Moon. In the science phase, the DARE satellite passes through the Earth-quiet cone on every orbit and simultaneously through the Sun shadow for 0 to 100% of every pass (Prime Science observations). Figure 1 illustrates DARE's observation conditions.

DARE's lunar science orbit is an outgrowth of the broad analytical and operational experience gained from the design and maintenance of LADEE's (the Lunar Atmosphere and Dust Environment Explorer) tightly controlled low-altitude and low-inclination lunar orbit, which achieved sustained 15-50 km periapsis altitudes over the sunrise terminator[3]. For DARE, a uniquely suitable orbit has been identified, consisting of a 50x125 km altitude "frozen" orbit with 0-3° inclination. Key to this orbit is a stable periapsis initially set at a Moon-fixed longitude of 180° which remains fixed over the lunar farside. DARE's frozen periapsis longitude of 180±20° maximizes Prime Science time and precludes the need for orbit maintenance maneuvers throughout the 2-yr

---

[*] Systems Engineer V, Metis Technology Solutions, NASA/Ames Research Center, Mission Design Center Moffett Field, CA.
[†] Aerospace Engineer, Ames Associate, NASA/Ames Research Center, Moffett Field, CA.
[‡] Professor, Center for Astrophysics and Space Astronomy, Department of Astrophysical and Planetary Sciences, University of Colorado Boulder

life of the mission. The stability of DARE's science orbit has been verified for all values of lunar orbit insertion RAAN and for inclinations as high as 3.5° using the GRAIL660 gravity model with degree & order of 100x100 and including third body effects.

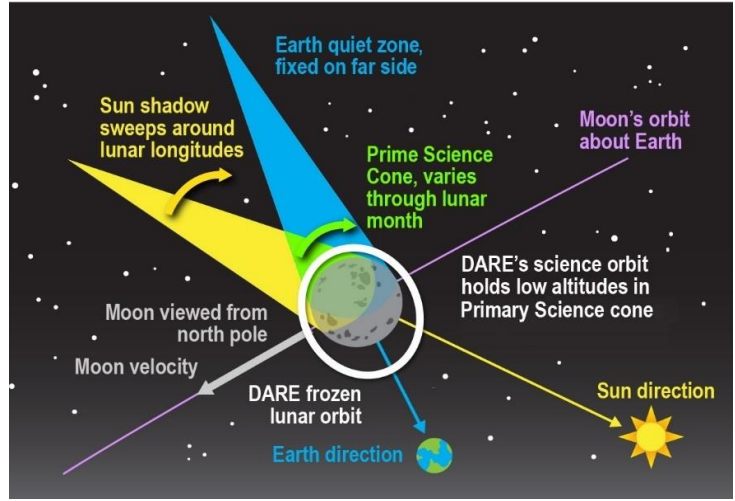

**Figure 1. Illustration of DARE's Lunar Orbit Depicting the Prime Science Observation Region at the Intersection of Earth and Sun EMI Quiet Cones.**

The DARE frozen orbit's stability appears in timeline and polar, e-omega formats in Figure 2 for the nominal duration of the DARE mission. The orbit is stable indefinitely, with long term, high fidelity propagations showing no discernible secular trend in the relevant mission parameters of apoapsis and periapsis altitudes and longitude of periapsis.

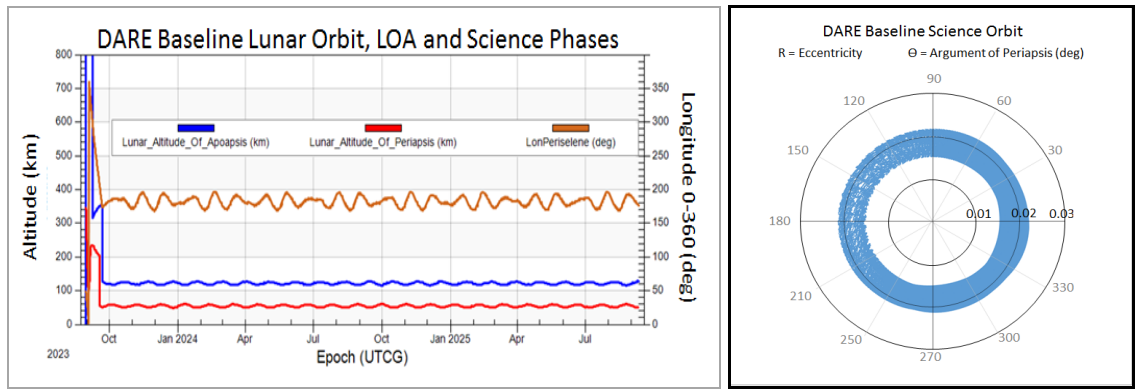

**Figure 2. The 2-year DARE lunar orbit with 50 x 125 km altitudes, inclination 1.72 degrees, and longitude of periselene frozen near 180 degrees. A) Timeline of LOA and science phases. B) Polar plot of eccentricity and argument of periapsis.**

Another key aspect of the DARE mission design is the science observation pointing strategy, which is constrained by the number of sky regions that can be accessed over the course of the mission, a minimum integration time associated with each target, and various pointing angle limits. As the Sun-Moon line traverses $360^0$ of right ascension, the Prime Science cone orients to different areas of the sky during the full moon throughout the year. Science targets have been identified and an observation timeline defined that take advantage of Prime Science opportunities while satisfying attitude constraints. Figure 3 depicts the targets on the Celestial Sphere, selected to be evenly distributed for viewing during the full moon configuration and to stay clear of the galactic center. Details on target selection and pointing constraints are given below under Attitude Design.

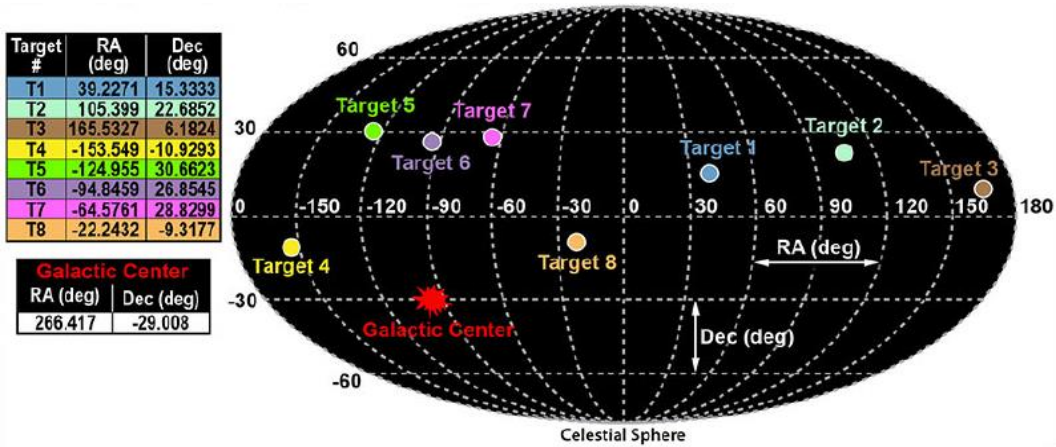

**Figure 3. DARE's Eight Observation Targets Depicted on the Celestial Sphere, with the Galactic Center Shown in Red.**

Placement of DARE's orbit periselene near the Earth-Moon line maximizes the time the spacecraft spends in the Prime Science cone. Figure 23 contains a graph of the cumulative hours of Prime Science (left axis) collected over the course of DARE's two year science mission.

The combination of DARE's frozen periselene on the lunar far-side and instrument observation strategy, enables DARE to exceed its science integration time requirement of 800 hours over a two year science mission and satisfy all pointing constraints. In fact, DARE's unique 50x125 km frozen orbit provides a total of 1150 Prime Science observation hours, an 18% increase in science hours over previous designs relying on a 125 km circular lunar science orbit[4].

**DARE TRAJECTORY DESIGN**

DARE mission design leverages heritage from the only other mission with a sustained low, near-equatorial lunar orbit: the LADEE mission, designed, built and operated by NASA's Ames Research Center[3]. DARE mission design borrows LADEE's phasing loops for lunar transfer[5], with an added characteristic of plane change between lunar orbit insertion maneuvers to establish equatorial inclination. The Phasing loop design accommodates $28.5^0$ or $37.5^0$ latitude for launch (i.e., Kennedy or Wallops, respectively) and offers two transfer planes per month: "Coplanar," near the Moon's orbit plane, and "Out of Plane (OOP)," with larger relative inclination.

After phasing loop transfer to the Moon, the first Lunar Orbit Insertion (LOI1) maneuver captures into a prograde orbit with 18 hour period. The post-capture orbit period is an effective balance of the drivers to mitigate Earth 3$^{rd}$ body perturbations and to perform the plane change maneuver at high altitude. B-theta targeting to date uses $2.5^0$ with the coplanar transfer type and $10^0$ with the OOP type. At lunar arrival, coplanar transfer trajectories have lower inclinations than OOP cases, but larger arguments of periapsis; design effort selected b-plane targeting to create affordable plane change maneuvers at high altitude. Figure 4 illustrates the DARE trajectory for all mission phases up to early Science Phase.

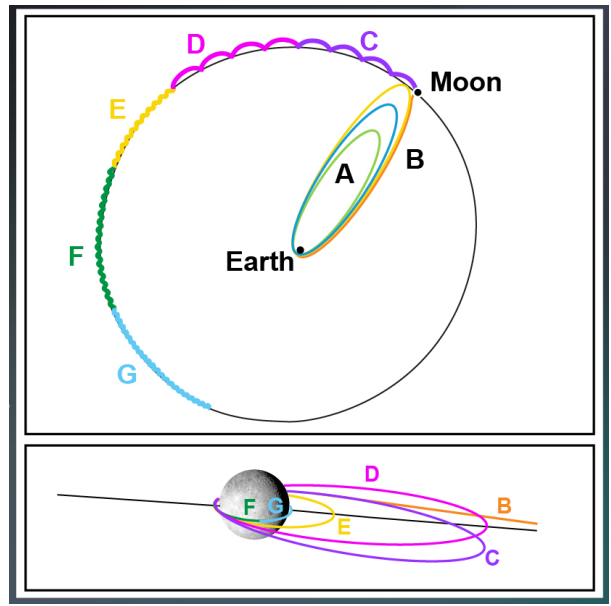

**Figure 4.** DARE trajectory showing phasing orbits (A), lunar transfer (B), lunar insertion 18 hr orbit (C), inclination trim from 5 or 20 deg to equatorial 0 - 3 deg (D), apoapsis lowering into 4 hr orbit (E), staging orbit (F), and frozen science orbit at 50 x 125 km (G).

**Table 1 - DARE Trajectory Maneuvers.**

| Mnvr # | Approx. Days from Launch | Days from Previous Maneuver | Maneuver Type | Nominal Magnitude or Allocation (m/s) | Description |
|---|---|---|---|---|---|
| -- | <0.1 | N/A | N/A | N/A | LV Phasing Orbit Injection & Separation |
| 1 | 3 | N/A | Engineering | 10 | AM1 Calibration of propulsion subsystem |
| 2 | 6 | 3 | Deterministic | 31 | PM1 Apo raising & phasing* |
| 3 | 9 | 3 | Deterministic | <10 | PAM (Plane change to compensate for launch window) |
| 4 | 14 | 5 | Deterministic | 3 | PM2 Apogee raising & phasing* |
| 5 | 24 | 10 | Deterministic | 0 | PM3 Apogee raising* |
| 6 | 25 | 1 | Statistical | <5 | TCM1 Correction |
| 7 | 27 | 2 | Statistical | <5 | TCM2 Correction |
| 8 | 29 | 2 | Deterministic | 304 | LOI1 capture into 18 hour orbit |
| 9 | 32 | 3 | Deterministic | 14 | LIM (Lunar Inclination Maneuver) |
| 10 | 35 | 3 | Deterministic | 269 | LOI2 lowers aposelene (4 hr. orbit) |
| 11 | 38 | 3 | Deterministic | 227 | LOI3 lowers aposelene |
| 12 | 48 | 9 | Deterministic | 31 | OLM1 Orbit Lowering Maneuver - Periselene |
| 13 | 51 | 3 | Deterministic | 47 | OLM2 Lower Aposelene |
| 14 | Variable | Variable | Deterministic | <7 | RFI avoidance if needed |
| 15 | 780 | Variable | Deterministic | 10 | Deorbit |
| | | | | | *Varies with launch day |

The DARE delta-v budget for the maneuvers in Table 1 is grounded on a comprehensive collection of analysis conducted for the LADEE mission, including year-long launch surveys and Monte Carlo simulations capturing all possible launch geometry and 3-σ dispersions[5]. The delta-v budget for the maneuvers in Table 1 uses allocations that accommodate worst case or 3σ simulation results. Using the maximum impulsive ΔV of 1150 m/s and allowing for 28 m/s for losses & statistical corrections, a total ΔV budget of 1260 m/s is allocated for DARE to acquire the lunar science orbit assuming a launch C3 of -2.78 $km^2/sec^2$, providing nearly 300% margin on statistical losses/ corrections. The Ball Aerospace Technology Corporation designed spacecraft has a wet mass of 1002 kg.

Third body gravitational perturbations (namely Earth) during Lunar Orbit Acquisition (LOA) phase cause lowering of periselene altitude, so the trajectory design establishes insertion altitude to provide 100 km of margin above the surface. Figure 5 shows a one year survey of lunar capture cases. The figure includes only periselene altitude for the LOA phase, covering 3 Lunar Orbit Insertion (LOI) maneuvers and the Lunar Inclination Maneuver (LIM) plane change. The two transfer planes, Coplanar and Out of Plane, appear in cool and warm colors, respectively. The four LOA maneuvers appear in Figure 5 on approximately 3 day intervals, selected for operational considerations

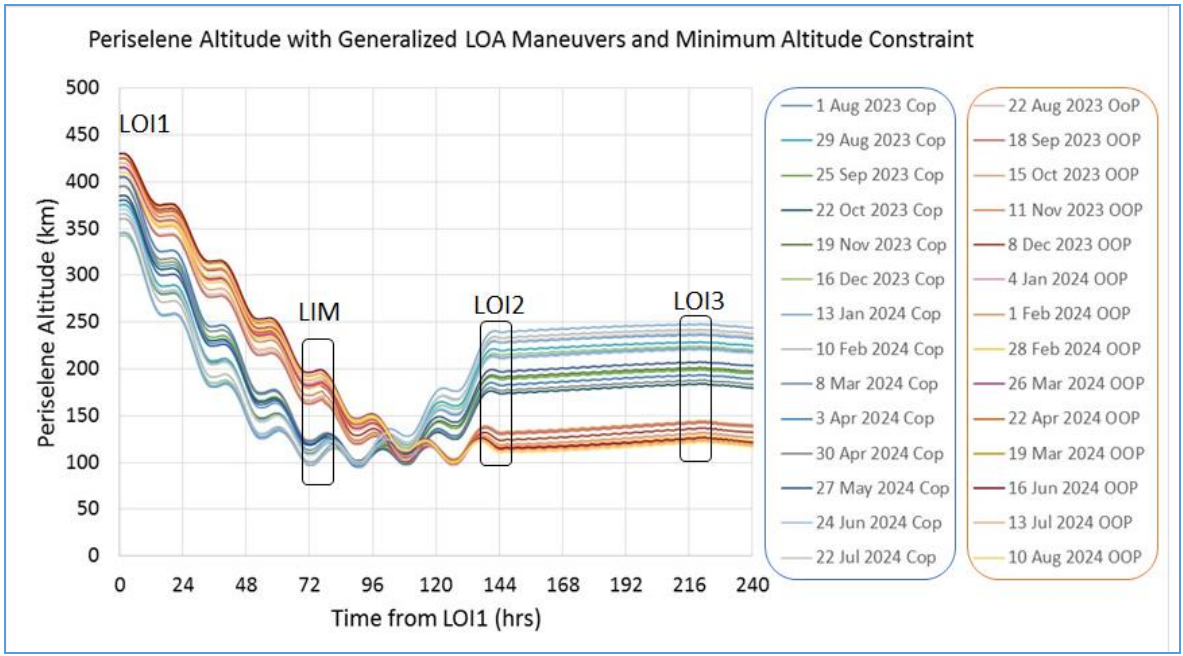

**Figure 5. DARE Lunar Orbit Acquisition Phase for One Year Survey of Trajectory Design Cases.**

Orbit transfer maneuvers to reach the frozen condition require management of three parameters: altitudes of both apsides and the alignment of the line of apsides. For operational simplicity and consistency, the DARE lunar orbit design uses a classic Hohmann transfer of in-track maneuvers for changing altitudes and coordinates the timing so that the altitude and alignment con-ditions are achieved at the right moment of the rotation of the Moon for the 180⁰ longitude of periselene. Figure 6 illustrates the orbit lowering approach for a one year survey of design cases. Figure 6 data curves are longitude of periselene in greens and grays, aposelene altitude in oranges and yellows, and in blue the continuations of the same periselene altitudes shown in Figure 5. The timing of the Orbit Lowering Maneuvers (OLM) is based on the trend of longitude of periselene toward 180⁰, with a three day minimum interval between maneuvers.

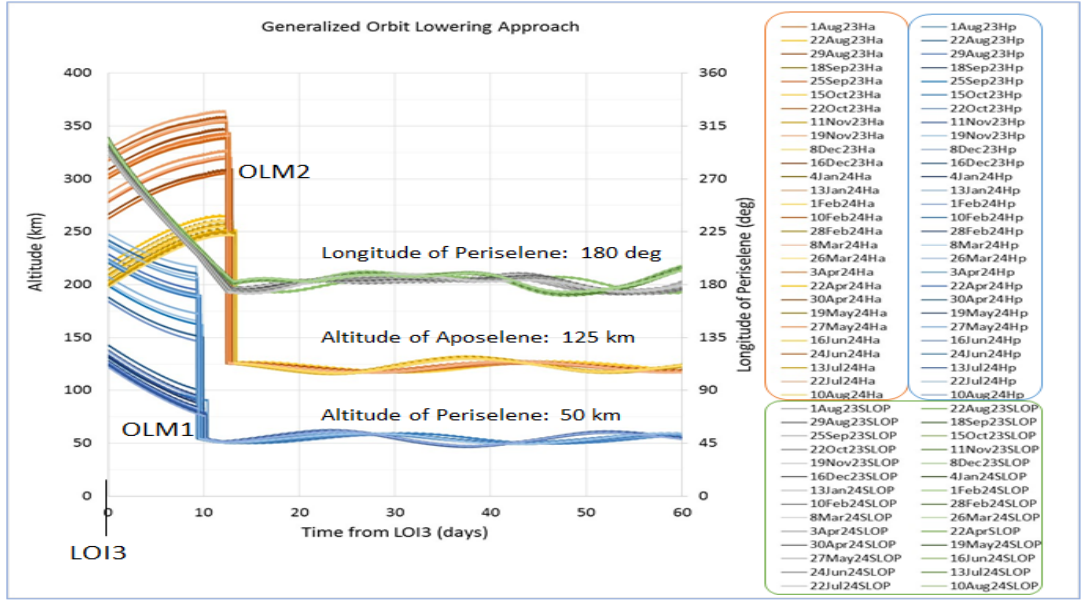

**Figure 6. Orbit Lowering Approach Achieves the Frozen Condition by Timing of the Maneuvers.**

**DARE LUNAR FROZEN ORBIT**

Low lunar orbits that are frozen with respect to lunar mascons are rare. Folta & Quinn[6] describe cases with altitudes similar to the DARE design but at higher inclinations. More research focuses on lunar orbits frozen with respect to third body perturbations, e.g. Ely & Lieb[7], while at very low altitude, lunar gravitational harmonics predominate.

Analytical treatments of lunar gravitational potential, to avoid becoming unwieldy, often include the harmonic terms up to J7 and fix the argument of periselene at 90⁰ or 270⁰ [8,9,6]. Results offer combinations of a, e, i, and ω to pursue with greater fidelity using computational tools.

The frozen condition in a low, equatorial orbit (Figure 2 above) resulted from computational analysis using gravitational terms at order and degree 100. The configuration is compatible with analytical solutions, but does not conform to typical constraints, particularly the fixed argument of periapsis. In the DARE orbit, the RAAN precesses relatively slowly (approx. 2/day) and the fluctuations in the longitude of periselene are minor, leaving argument of periapsis to revolve with each lunar cycle (Figure 2b).

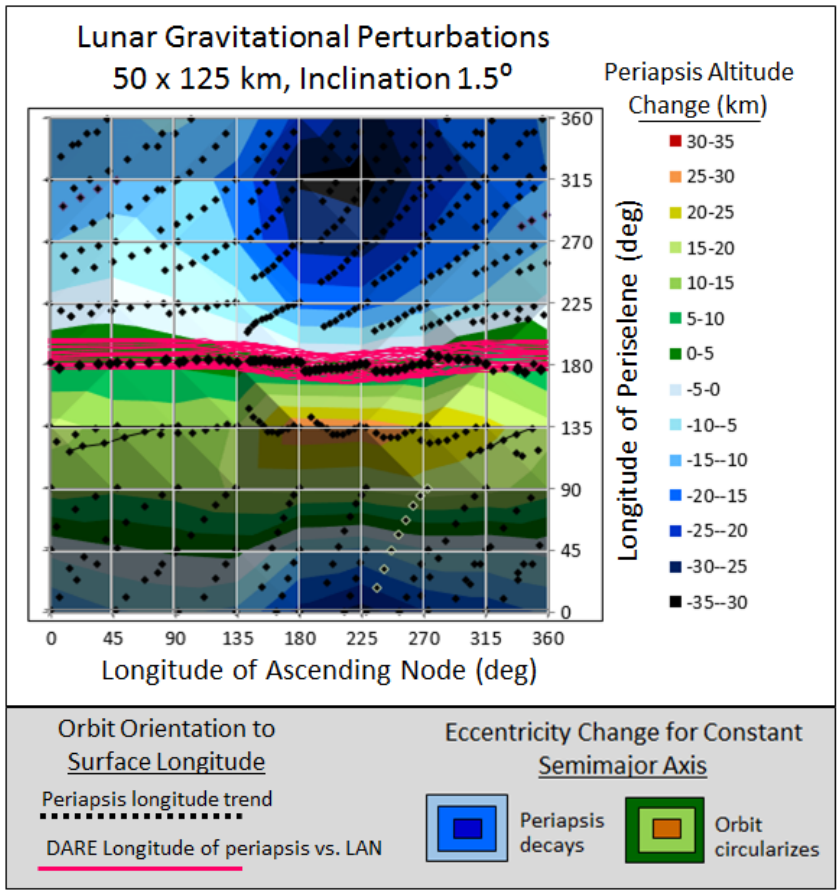

**Figure 7. DARE Lunar Orbit in the Conceptual Design Space Showing Trends in Altitude and Orientation.**

Plice & Craychee introduced a "contour map" method for characterizing lunar gravitational perturbations in low, equatorial, lunar orbit[10]. Figure 7 portrays the DARE baseline science orbit in the design space defined by longitude of periselene vs. longitude of ascending node. The x-axis shows the lunar month in approximately 3.5 day increments as the Moon's rotation continuously decreases the longitude of equatorial crossing, while the y-axis represents placement of the line of apsides, also referenced to selenographic longitude. For near-circular, equatorial orbits increments of periselene longitude are equivalent to increments of the argument of periapsis. Higher inclination orbits demonstrate similar patterns using the argument of periapsis parameter[10,11]. Background colors, calculated at 8x8 grid points, indicate the change in periselene altitude from that set of initial conditions during the time interval until the node longitude reaches the next x-axis grid line. Dotted, black, right-to-left trend lines show the change in longitude of periselene during the same interval, again from initial conditions at grid points.

Data lines of longitude of ascending node vs. longitude of periselene for the two-year DARE mission appear in a narrow region centered on the lunar far side (magenta), underscoring the stability of the line of apsides. Black trend lines for 45° increments of the Moon's rotation (x-axis grid) create visual guidelines for perturbation effects, while the DARE trajectory follows similar patterns, not limited to grid approximations.

Subsections below explore the scope of available frozen cases by varying orbital elements. Data plots show two years as baselined in the DARE science phase, unless otherwise specified.

**Semi-major Axis,** *a*

Varying semi-major axis while holding eccentricity fixed yields additional viable candidate frozen orbits (Figure 8), albeit with minor increase in the excursions in periapsis altitude and longitude. For DARE parametric trade studies, lower periselene altitudes provide longer Prime Science passes, at the expense of slightly greater lunar limb incursions during observations (see discussion on Attitude Design), and slightly higher operational risk.

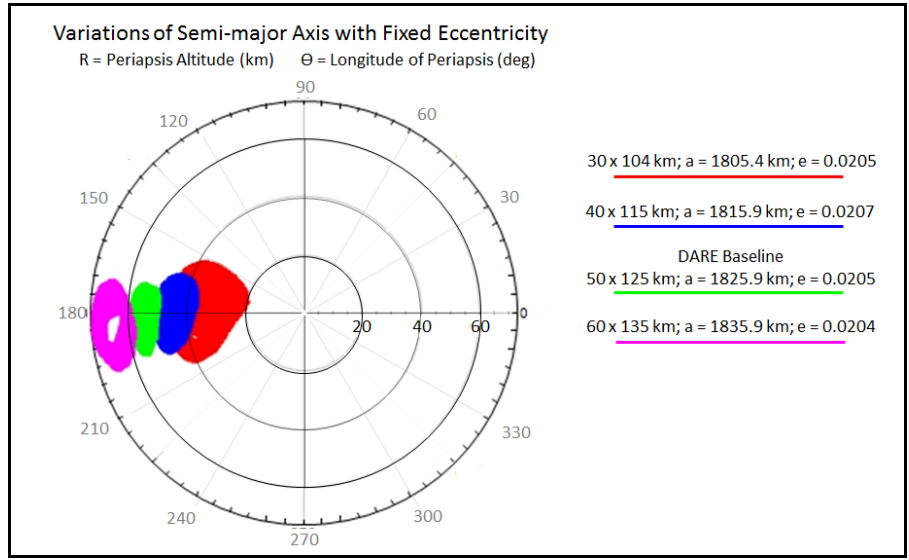

**Figure 8. Periapsis Altitude vs. Longitude for Variations of Initial Semi-major Axis.**

**Eccentricity,** *e*

The frozen orbit is more sensitive to eccentricity than semi-major axis. Figure 9 compares the DARE baseline to other values of eccentricity. Periapsis stability shows larger cycles of longitude than in the set of cases varying semi-major axis in Figure 8.

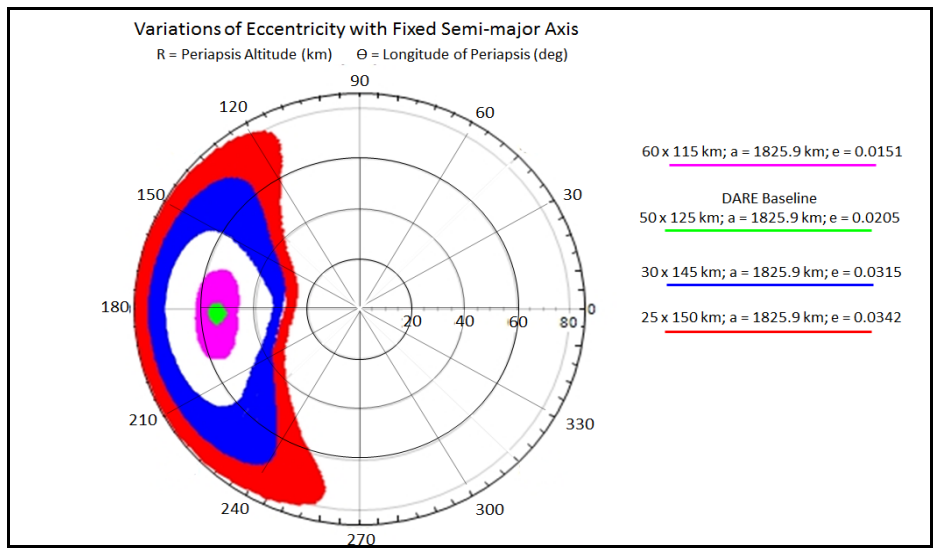

**Figure 9. Periapsis Altitude vs. Longitude for Variations of Eccentricity.**

**Inclination, *i***

Figure 10 explores the range of viable inclinations for the frozen orbit. Altitude and orientation stability begin to degrade at inclinations greater than about $3.5°$. Pure equatorial orbits demonstrate stability, but introduce modeling inconvenience with undefined node. Inclination of $0.5°$ is a good representation of equatorial orbit.

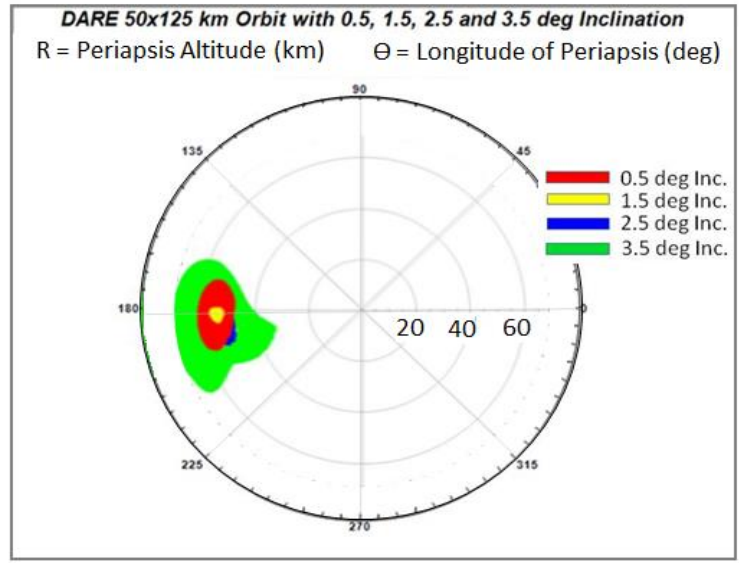

**Figure 10. Orbit Stability near the Lunar Equator.**

Frozen equatorial orbits show regular cycles of approximately $1.5°$ excursions in the value of inclination. Figure 11 illustrates the pattern of oscillations in inclination for two years of lunar orbit at initial inclinations from $0.5°$ to $3°$.

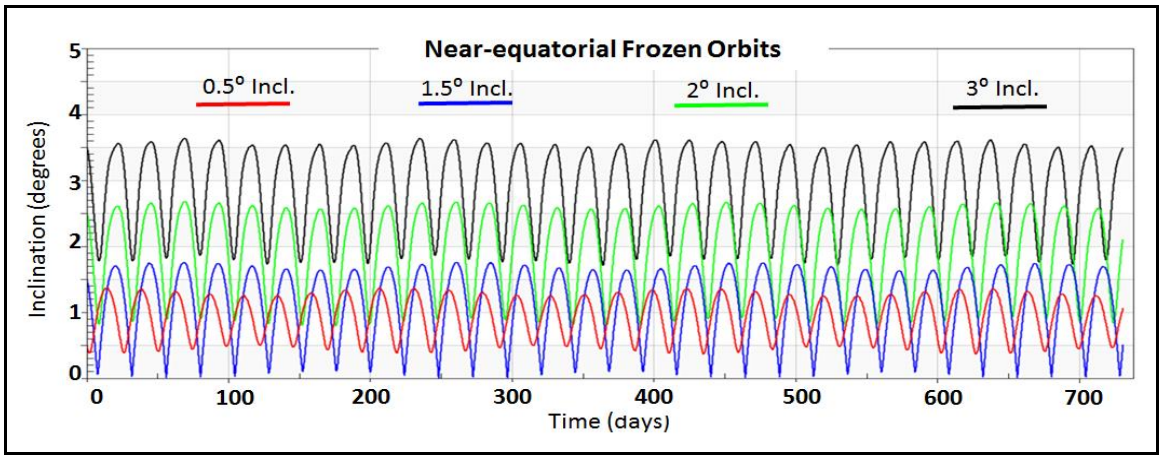

**Figure 11. Inclination Varies by Approximately $1.5°$ in Frozen, Equatorial Lunar Orbit.**

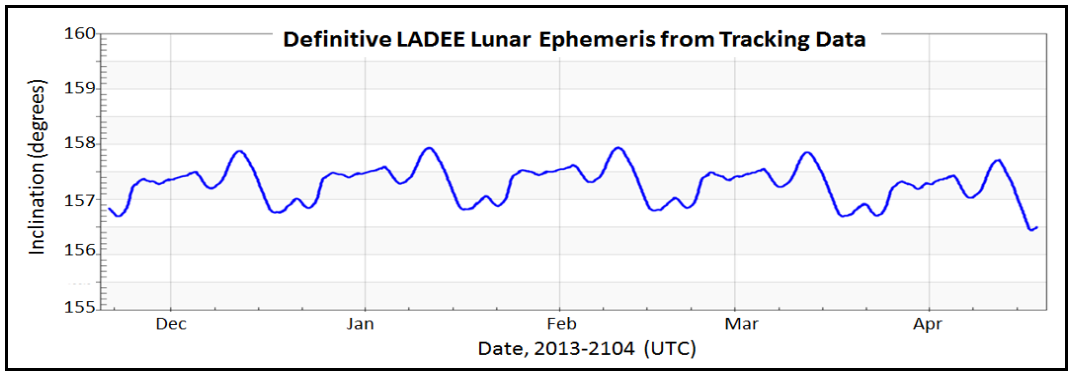

**Figure 12. LADEE Definitive Ephemeris Shows Regular, Small Fluctuations in Inclination.**

LADEE's approximately 5 month mission orbit was retrograde, near-equatorial at $157^0$ inclination (Figure 12). The parallel with DARE's frozen, equatorial orbit is further limited in that LADEE's orbit had sizeable changes in $a$ and $e^{3,10}$ from maneuvers and gravitational perturbations, which create the "sawtooth" pattern in each lunation in Figure 12. The DARE and LADEE lunar orbits show similar patterns of inclination change in each lunar cycle, varying approximately $1.5^0$.

**Right Ascension of the Ascending Node, RAAN**

Direct and phasing loop transfer trajectories typically arrive at lunar capture in a relatively narrow range of right ascension and longitude of the ascending node, however non-traditional designs, such as weak stability boundary trajectories[4] may arrive at a wider variety of node locations. Once established, the frozen DARE orbit is stable for any value of RAAN. Figure 13 presents a polar plot of science orbit periapsis longitude and altitude over 24 months across all RAANs. In all cases, periapsis remains fixed over the lunar far side.

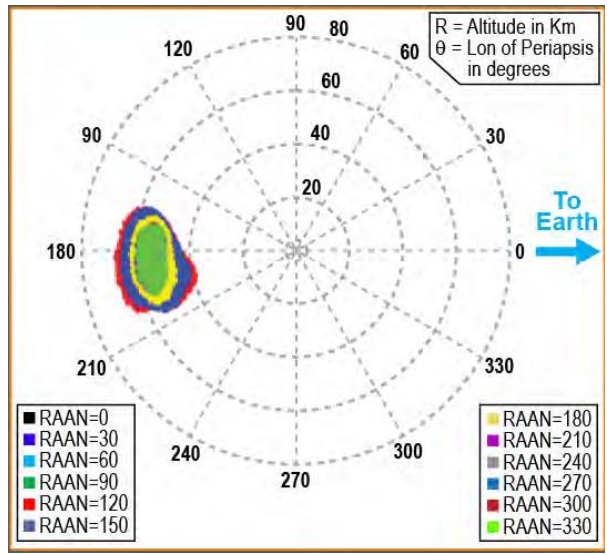

**Figure 13. Polar Plot of Science Orbit Periapsis Longitude vs. Altitude across all RAANs.**

**Longitude of Periapsis**

DARE science requirements call for maximizing operating time near lunar longitude 180⁰. Missions with other desired orientations may not find benefit from this frozen orbit approach; most initial values of periselene longitude offer no frozen characteristic and often decay to impact quickly, while Figure 7 shows that for longitudes between approximately 150⁰ and 220⁰, the gravitational perturbations create trends across lunar far side. Small offsets from the target 180⁰, such as those that would be present with typical maneuver errors or other uncertainties, fall within the stable condition.

The parametric relationship between longitude and altitude of periapsis shows repeated cycles in Figure 14, which could offer stability sufficient for some mission applications, however the traditional e-omega plots exhibit ongoing secular trends in Figure 15, significantly in contrast to the frozen orbit signature in Figure 2 and in Folta & Quinn, 2006[6].

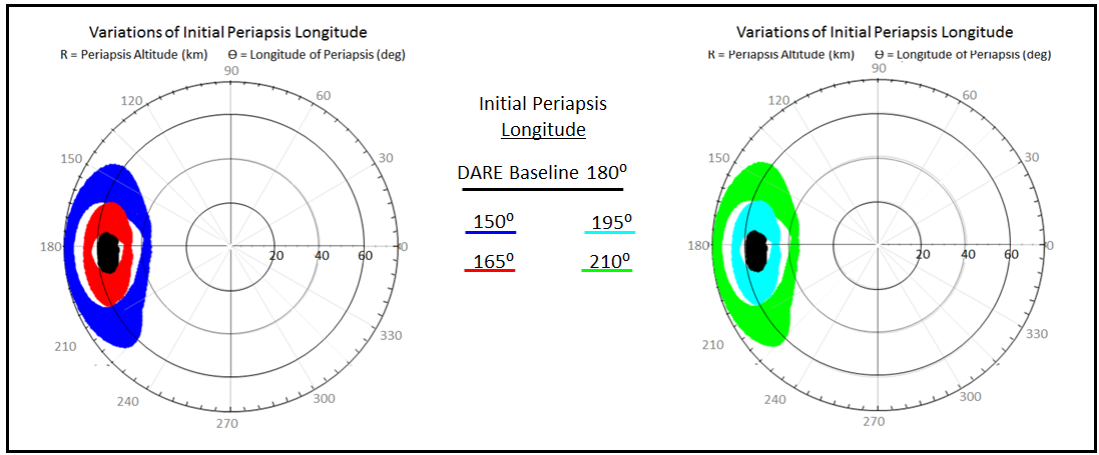

**Figure 14. Periapsis Altitude vs. Longitude for Variations of Initial Conditions.**

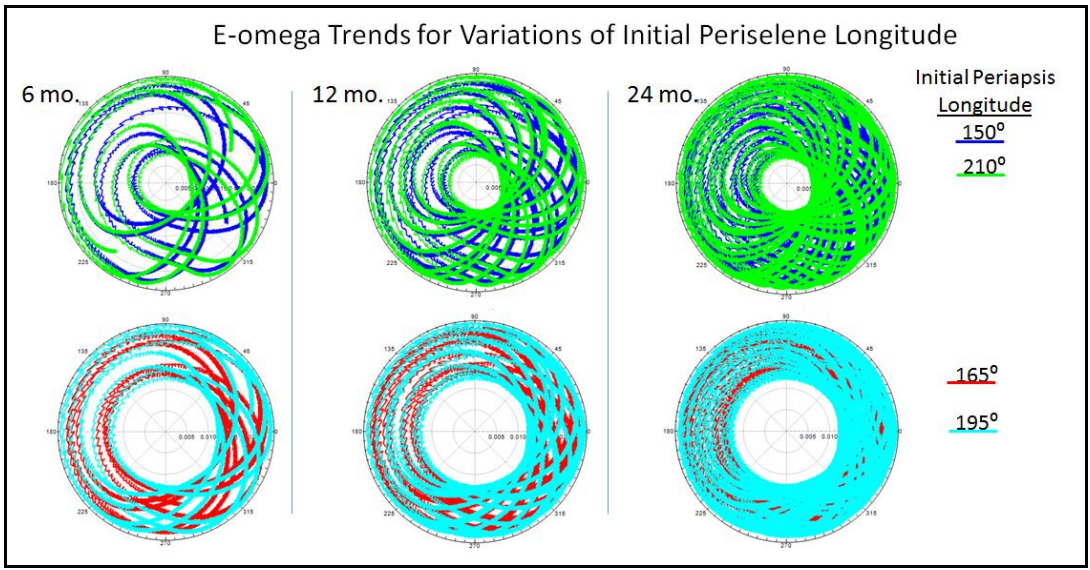

**Figure 15. E-Omega Plots for 6, 12, and 24 Month Durations.**

As expected, the cases with alternate values of initial periapsis longitude, but not far off the midpoint of the lunar far side, follow the same general trend of periselene longitude vs. node longitude as the baseline case, but with greater variation. Figure 16 illustrates the same five cases as Figure 14, this time in the contour map format, showing trends in time as the Moon rotates within the orbit. Color codes match Figure 14, though at the cost of some overlap with the standardized background colors which were borrowed from cartographic conventions for intuitive display.

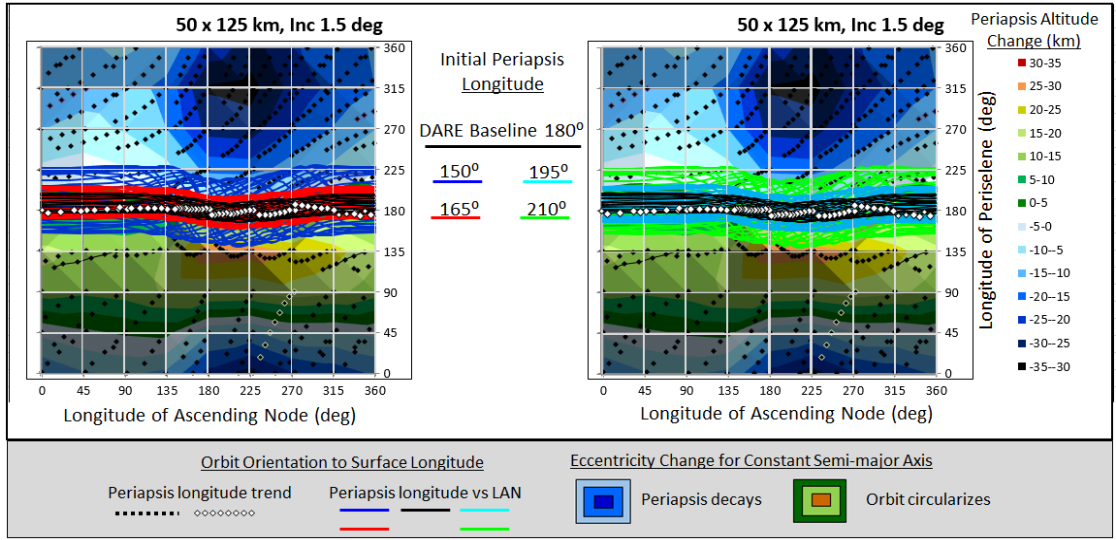

**Figure 16. Longitude of AN vs. Longitude of Periapsis "Contour Map" for Cases with Alternate Initial Values of Periapsis Longitude.**

**Argument of Periapsis, $\omega$**

Other studies have identified frozen orbit stability with argument of periapsis values of 90° and 270°.[6,9] The DARE baseline case has the initial line of nodes nearly in alignment with the line of apsides (argument of periapsis near zero). It is typical for orbits to be nearly fixed inertially, with some precession of the line of nodes. For a closely controlled longitude of periapsis, it is consistent that the longitude of the ascending node (LAN) and argument of periapsis (AoP) would cycle regularly with the Moon's rotation. Figure 17 illustrates precession of the line of nodes roughly twice per year, with cycles of LAN and AoP with each lunation for representative values of initial AoP: zero (DARE baseline), 90, 180, & 270 degrees. Longitude of periapsis stays near 180° degrees in all cases. Figure 17 and Figure 18 display data for timespans of six months or one year, for greater clarity among cases.

There is a configuration that appears to offer extra stability, at the initial value of 210° LAN (AoP 330° for maintaining longitude of periselene at 180°). For a very limited set of cases, the longitude of the ascending node and argument of periapsis hold to narrow ranges while the RAAN cycles with each lunation. While the parameters never appear to approach being poorly defined due to low inclination or low eccentricity, the low value of inclination probably accounts for the ability of the RAAN to cycle more rapidly than normal precession.

Figure 18 shows examples of unusual cases with extra stability, with both the line of apsides and the line of nodes frozen with respect to lunar longitude. The case with greatest stability identified to date has LAN = 210°, $a$ = 1821 km, $e$ = 0.0178, $i$ = 0.5°, and longitude of periapsis 180°.

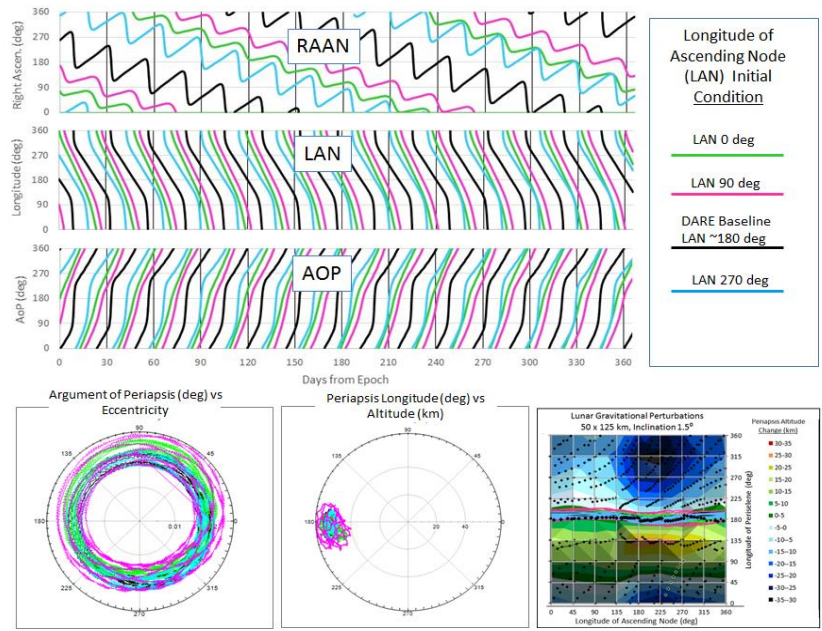

**Figure 17. Combinations of Parameters for Frozen Orbits: RAAN vs. Time, LAN vs. Time, AoP vs. Time, E-Omega, Periapsis Longitude & Altitude, and Contour Map.**

The version of the contour map in Figure 18 has custom features to illustrate the trends of longitude of periselene vs. longitude of ascending node. Background colors indicate decay trends standardized to the same durations as Figure 7 and Figure 17, with customized time intervals for trends in gravitational perturbations selected for clarity.

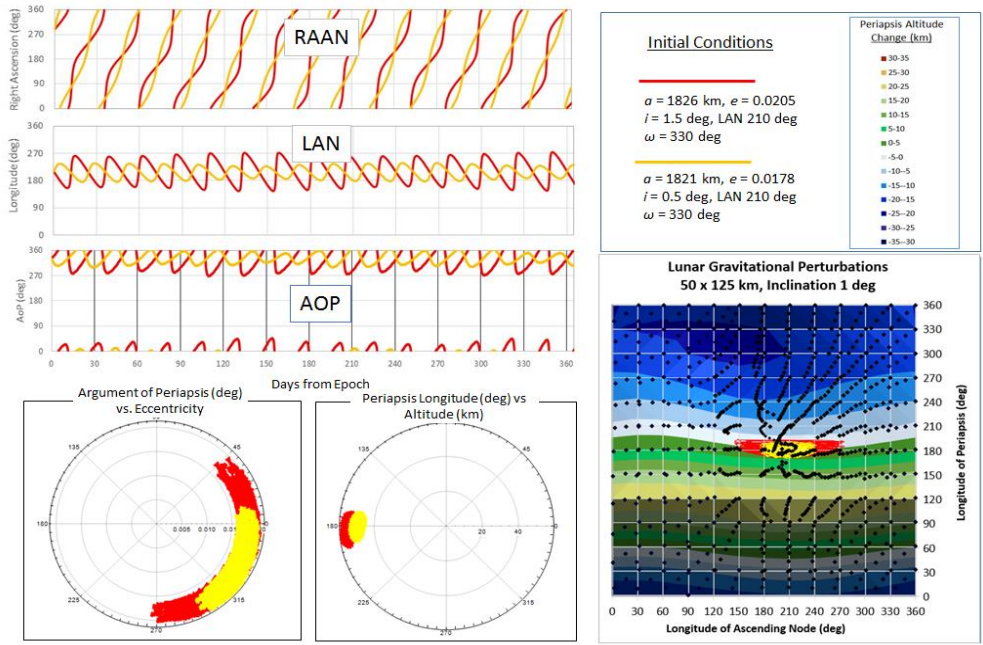

**Figure 18. Combinations of Parameters for Frozen Orbits with Unusual Stability: RAAN vs. Time, LAN vs. Time, AoP vs. Time, E-Omega, Periapsis Longitude & Altitude, and Contour Map.**

**Summary of DARE Lunar Orbit Design**

For the right combination of mission requirements, frozen orbits may offer design solutions with the line of apsides oriented to lunar longitude. A significant advantage of frozen orbits is the elimination of maintenance maneuvers. In contrast to the DARE mission requirements for longitude placement only, the LADEE mission orbit was oriented to the sunrise terminator, requiring substantial design and operational effort in orbit maintenance, with maneuvers sometimes as frequently as 4 days apart[3,10].

The DARE baseline orbit presented above has semi-major axis of 1825.9 km, eccentricity of 0.0205, inclination of 1.5°, and longitude of periselene at 180°. For applications that can accommodate wider fluctuations of longitude alignment, frozen orbit solutions exist within ranges of parameters listed in Table 2. Also included are input values for greater stability.

**Table 2. Summary of Frozen Orbit Parameters.**

| Parameter | DARE Baseline | Wider Constraints | Tighter Constraints |
|---|---|---|---|
| Semi-major Axis (km) | 1825.9 | 1805.4 – 1835.9 | 1820.9 – 1835.9 |
| Eccentricity (deg) | 0.0205 | 0.0151 – 0.0342 | 0.0178 – 0.0205 |
| Inclination (deg) | 1.5 | 0.5 – 3.5 | 0.5 – 1.5 |
| Longitude of Periapsis (deg) | 180 | 150 - 210 | 180 |
| Initial Longitude of AN (deg) | 180 | 0 - 360 | 210 |

**DARE ATTITUDE DESIGN**

DARE's attitude design seeks to satisfy science pointing requirements that entail long integration time stares at fixed sky regions over portions of the orbit that are shielded from Earth and solar radio emissions. Table 3 lists DARE's observation requirements and constraints, which include limits on how close the instrument/antenna boresight can be pointed to the lunar limb and to the galactic center during observations; a limit on the total number of allowable science observation targets over the course of the mission; a minimum number of integration time hours associated with each target; and a minimum total Prime Science observation hours over the course of the two year science mission.

**Table 3. DARE Prime Science Requirements/Constraints during RFI Quiet Observations.**

| Requirement/Constraint | Value |
|---|---|
| Moon Limb Angle Limit | 40 deg |
| Galactic Angle Limit | 55 deg |
| Minimum # of Science Targets | 4 |
| Maximum # of Science Targets | 8 |
| Minimum Science Observation Integration Time per Target | 50 hours |
| Total Science Integration Time | 800 hours |

DARE Prime Science observations must occur during periods when the spacecraft traverses through RFI quiet regions created by lunar occultation of the Earth and Sun. Such occultations produce conical regions during which observations can take place. The apex of the Earth RFI quiet cone is aligned with the Earth-Moon line and is situated at the far side of the Moon (Figure 19), such that it rotates 360 degrees over the course of a sidereal month. The apex of the Sun quiet cone is aligned with the Sun-Moon line, such that it rotates 360 deg over the course of a year, relative to inertial space. Prime Science opportunities occur every month during passages through the region formed by intersection of the two cones. This intersection is at its maximum during full moon, when Earth/Sun RFI-free observations reach a peak duration of 28 minutes per orbit, and gradually decrease to zero between last and first quarter lunar phases (Figure 20).

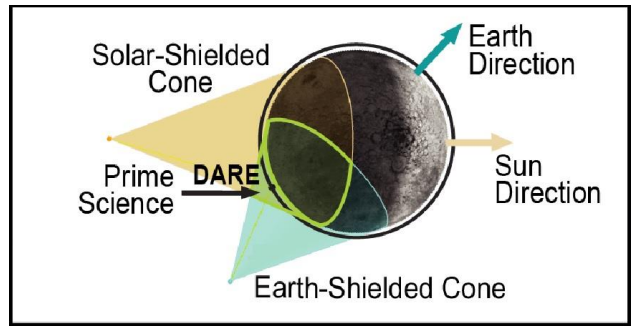

**Figure 19. DARE Prime Science Observation Region Formed by the Intersection of RFI Earth-Quiet and Sun-Quiet Cones.**

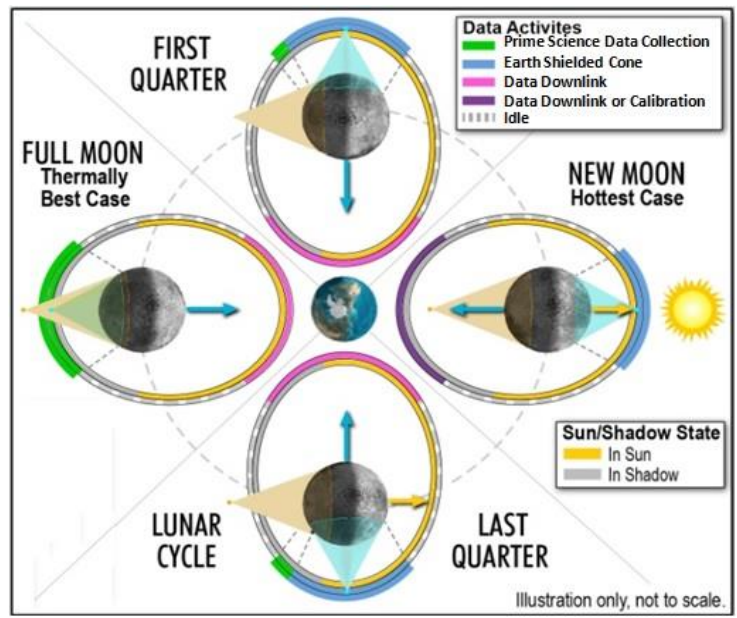

**Figure 20: DARE Prime Science Observation Geometry Provides a Maximum of 28 minutes of RFI-Free Observations Each Orbit at Full Moon.**

In order to satisfy the instrument Moon limb angle constraint during observations, one would ideally select the Earth-Moon vector associated with each full moon event as a target, for a total of twelve separate inertial science pointing targets over the course of a year. However, a con-straint limiting the total number of science targets to no more than 8 throughout the mission (Table 3) necessitates the re-use of targets across full moon events. Furthermore, an additional constraint placed on the minimum allowable angle between the instrument/antenna boresight and the galactic center requires the use of high ecliptic elevation targets at certain times of the year (when the Sun-Earth vector approaches the galactic center). These factors impose limits on the size of the lunar limb angle that can be achieved during DARE prime science observations, and require a trade against the height of the instrument sun shade needed to block the lunar albedo from reaching the antenna, in order to satisfy instrument thermal requirements.

Table 4 lists the eight preliminary DARE observation targets and the observation timeline defined over its 2-year science mission that spans a total of 25 lunar synodical months. Targets 5, 6, and 7 consist of the high elevation targets chosen to avoid the galactic center.

**Table 4: DARE Science Targets and Observation Timeline over a Two Year Science Mission, Assuming an August 1, 2023 Launch Date.**

| Full Moon Event | Full Moon Epoch | Adjusted Target RA (deg) | Adjusted Target Dec (deg) | Consolidated Target RA (deg) | Consolidated Target Dec (deg) | Target # | Start Epoch |
|---|---|---|---|---|---|---|---|
| 1 | 28 Oct 2023 20:24:05.05 | 32.55 | 13.13 | 39.2271 | 15.3333 | 1 | 15 Oct 2023 00:00:00.00 |
| 2 | 27 Nov 2023 09:16:20.20 | 62.55 | 21.04 | | | | |
| 3 | 27 Dec 2023 00:33:15.15 | 95.05 | 23.36 | 105.399 | 22.6852 | 2 | 15 Dec 2023 00:00:00.00 |
| 4 | 25 Jan 2024 17:54:04.04 | 127.26 | 19.04 | | | | |
| 5 | 24 Feb 2024 12:30:30.30 | 156.88 | 9.66 | 165.5327 | 6.1824 | 3 | 10 Feb 2024 00:00:00.00 |
| 6 | 25 Mar 2024 07:00:22.22 | -175.61 | -1.90 | | | | |
| 7 | 23 Apr 2024 23:49:00.00 | -148.29 | -12.84 | -153.549 | -10.9293 | 4 | 10 Apr 2024 00:00:00.00 |
| 8 | 23 May 2024 13:53:10.10 | -119.49 | 29.32 | -124.955 | 30.6623 | 5 | 10 May 2024 00:00:00.00 |
| 9 | 22 Jun 2024 01:07:55.55 | -89.15 | 26.56 | -94.8459 | 26.8545 | 6 | 10 Jun 2024 00:00:00.00 |
| 10 | 21 Jul 2024 10:17:12.12 | -59.07 | 29.60 | -64.5761 | 28.8299 | 7 | 10 Jul 2024 00:00:00.00 |
| 11 | 19 Aug 2024 18:25:52.52 | -30.87 | -12.54 | -22.2432 | -9.3177 | 8 | 05 Aug 2024 00:00:00.00 |
| 12 | 18 Sep 2024 02:34:31.31 | -4.28 | -1.85 | | | | |
| 13 | 17 Oct 2024 11:26:26.26 | 22.44 | 9.40 | 39.2271 | 15.3333 | 1 | 05 Oct 2024 00:00:00.00 |
| 14 | 15 Nov 2024 21:28:32.32 | 51.28 | 18.69 | | | | |
| 15 | 15 Dec 2024 09:01:43.43 | 82.96 | 23.28 | 105.399 | 22.6852 | 2 | 01 Dec 2024 00:00:00.00 |
| 16 | 13 Jan 2025 22:26:57.57 | 115.51 | 21.37 | | | | |
| 17 | 12 Feb 2025 13:53:27.27 | 146.07 | 13.60 | 165.5327 | 6.1824 | 3 | 01 Feb 2025 00:00:00.00 |
| 18 | 14 Mar 2025 06:54:42.42 | 174.12 | 2.55 | | | | |
| 19 | 13 Apr 2025 00:22:17.17 | -158.74 | -8.93 | -153.549 | -10.9293 | 4 | 01 Apr 2025 00:00:00.00 |
| 20 | 12 May 2025 16:55:57.57 | -130.56 | 31.77 | -124.955 | 30.6623 | 5 | 01 May 2025 00:00:00.00 |
| 21 | 11 Jun 2025 07:43:52.52 | -100.56 | 26.92 | -94.8459 | 26.8545 | 6 | 25 May 2025 00:00:00.00 |
| 22 | 10 Jul 2025 20:36:51.51 | -69.99 | 27.84 | -64.5761 | 28.8299 | 7 | 25 Jun 2025 00:00:00.00 |
| 23 | 09 Aug 2025 07:55:07.07 | -40.91 | -15.85 | -22.2432 | -9.3177 | 8 | 25 Jul 2025 00:00:00.00 |
| 24 | 07 Sep 2025 18:08:56.56 | -13.80 | -5.90 | | | | |
| 25 | 07 Oct 2025 03:47:38.38 | 12.68 | 5.44 | 39.2271 | 15.3333 | 1 | 20 Sep 2025 00:00:00.00 |

For the step 1 proposal, a lunar limb angle constraint of 40° was established, which reduces the length of the sunshade extension beyond the antenna to 3 meters, with a trade study to be performed in Phase A to finalize the thermal shielding design. Given a 40 deg limb angle limit, a total of 1150 hours of Prime Science observations can be achieved over two years in the DARE science orbit, with each target accumulating over 80 hours of observation time (Figure 21 and Figure 24).

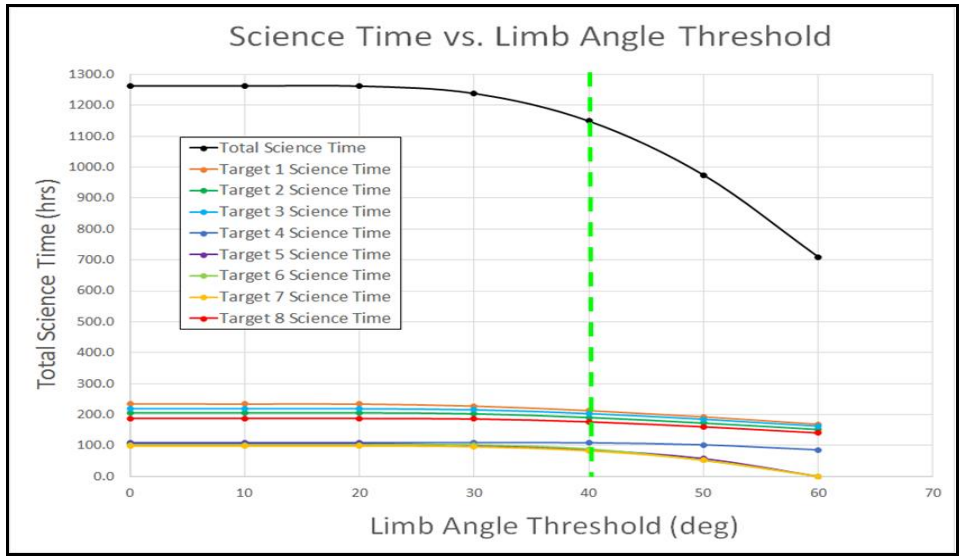

**Figure 21. DARE Prime Science Observation Time (Total and for Each Target) over Two Years as a Function of Lunar Limb Angle Constraint.**

Histograms of the duration of DARE's Prime Science observation passes in each orbit appear in Figure 22 for a single month and over the entire two-year science mission, indicating the availability of a relatively large number of long-duration passes. For practical reasons, the attitude design applies a cutoff of 2 minutes to eliminate short-duration passes, which reduces the overall number of passes by 6%, while only reducing total observation time by 5 hours (<0.5%) over the entire mission.

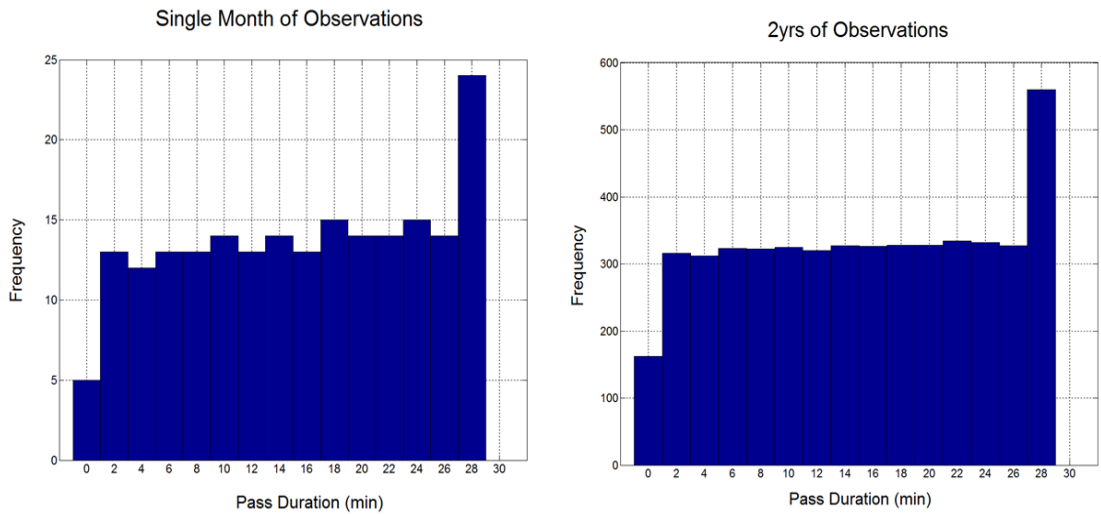

**Figure 22. DARE Prime Science Observation Histograms for a Single Month (left graph) and the Entire Two Year Science Mission (right graph).**

Given the combination of a frozen science orbit, which maximizes observation time in the Earth/Sun RFI quiet zone, and a set of target observation attitudes that maximizes antenna angles to the lunar limb and galactic center, the DARE mission design is able to satisfy all science requirements and constraints. Figure 23 contains a graph of the cumulative 1150 hours of Prime Science (left axis) collected over the course of DARE's two year science mission utilizing the observation timeline reflected in Table 4. Lunar limb angles relative to the instrument/antenna boresight which satisfy the $40^0$ minimum angle constraint are plotted on the right axis for each observation. The angle between the instrument/antenna boresight and the galactic center is also plotted on the right axis for each numbered target observation attitude (Figure 3), and satisfies a $55^0$ constraint.

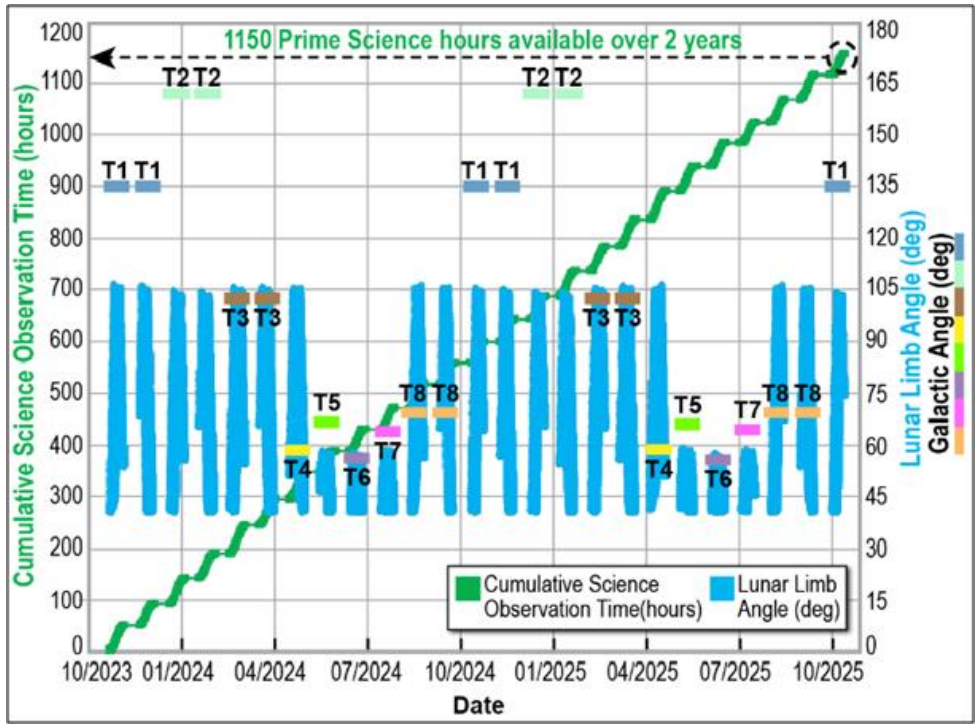

**Figure 23: DARE Cumulative Prime Science Hours (green curve, left axis) and Target Observation Timeline (with lunar limb and galactic angle, right axis).**

Relative to a requirement of 800 hours of Prime Science observation time, the 1150 hours provided by DARE's mission design yields a healthy 31% margin (Table 5). This margin also accounts for the possibility of up to 100 hours of interference between current or future lunar orbiting spacecraft.

The potential exists for RFI from lunar orbiting S/C that are within ~5000 km of DARE. An analysis performed to assess the amount of time when the Lunar Reconnaissance Orbiter (LRO) and the two aging THEMIS S/C (in separate ~500 x 16000 km orbits) could affect DARE Prime Science observations yielded a maximum of 169 hours of possible interference over 2 years (69 hrs for LRO and 100 hrs for THEMIS interference). However, simulations have shown that small infrequent DARE phasing maneuvers (0.5 m/s quarterly, for a total ΔV of 4 m/s over 2 years) can eliminate interference from LRO, or other future S/C in low lunar orbits. Given that the THEMIS spacecraft (both operating in ~24-hour period orbits) experience periapsis crossings where the spacecraft are within 5000 km of the Moon for over 2 hrs, it is unlikely that DARE phasing maneuvers could be effectively applied to avoid line-of-sight proximity passages. Thus, an allocation of 100 hours of possible RFI interference must be book-kept for THEMIS, which still yields a net of 250 hours of science observation margin for DARE (Table 5).

Table 5: Total DARE Prime Science Observation Time and Calculated Margin.

| | |
|---:|---:|
| Minimum Prime Science Time over 2 yrs | 1150 hrs |
| Required Prime Science Time | 800 hrs |
| Possible EMI due to THEMIS or other s/c | 100 hrs |
| Science Time Margin | 250 hrs |
| Percent Science Time Margin (250/800) | 31% |

## CONCLUSION

Custom orbit and attitude designs facilitate the Dark Ages Radio Experiment mission, currently proposed for NASA's Astrophysics MidEx program to use the radio quiet zone on the far side of the Moon to explore the cosmic dark ages. The translunar and lunar orbit insertion trajectory borrows advantages from the LADEE mission, while the novel, frozen lunar orbit maximizes time in science observation conditions indefinitely without maintenance maneuvers. Target regions of the sky, selected to optimize time in the radio quiet zone and avoid the galactic center provide robust observation times. Figure 24 illustrates the success of the design in exceeding the required 800 hours of science observation time.

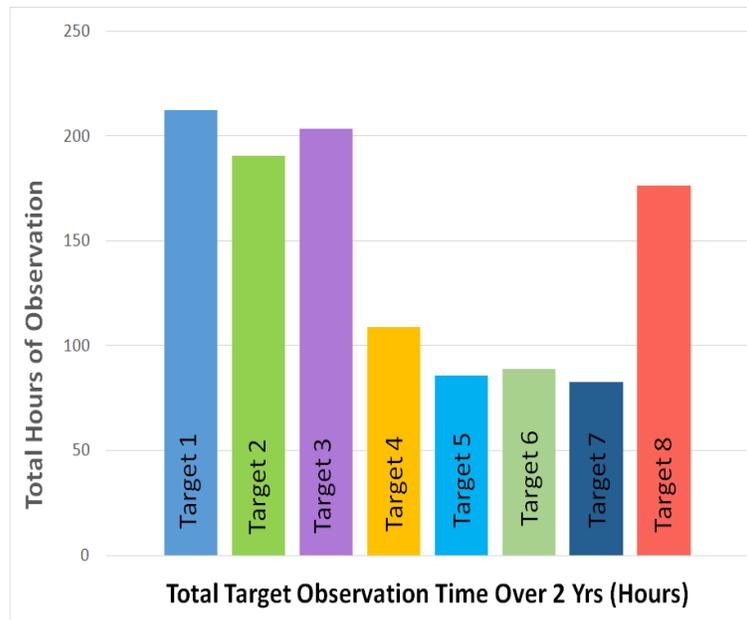

**Figure 24.** Total Observation Time by Target for the DARE Baseline Mission.

## ACKNOWLEDGMENTS

The authors would like to recognize Lisa Woloszyn for her extraordinary talent in creating technical images. Further appreciation goes to Joy Colucci, President of Metis Technology Solutions, and Scott Richey, Division Chief of the Ames Mission Design Division, for their support in the writing of this paper. As always, enduring gratitude goes to the late Professor Harm Buning, whose many students of orbital mechanics included astronauts from Projects Mercury, Gemini, and Apollo.